\documentclass[aps,10pt,twocolumn,superscriptaddress,footinbib,flushbottom,floatfix]{revtex4-1}%
\usepackage{amssymb,amsmath,amsfonts,bm,dcolumn}
\usepackage[symbol]{footmisc}
\usepackage{graphicx,color}
\usepackage[squaren]{SIunits}
\usepackage[english]{babel}
\usepackage{tabularx}
\usepackage{tikz}
\usepackage{epstopdf}
\usepackage{enumitem}

\usepackage{times}

\begin{document} 


\title{Janus microswimmers are poor hydrodynamic mixers}
%
	
	\author{Maximilian R. Bailey}
	\affiliation{Laboratory for Soft Materials and Interfaces, Department of Materials, ETH Z{\"u}rich, Vladimir-Prelog-Weg 5, 8093 Z{\"u}rich, Switzerland}
	
	\author{Dmitry A. Fedosov}
	\affiliation{Theoretical Physics of Living Matter, Institute of Biological Information Processing and Institute for Advanced Simulation, Forschungszentrum J{\"u}lich, 52425 J{\"u}lich, Germany}

        \author{Federico Paratore}
	\affiliation{Laboratory for Soft Materials and Interfaces, Department of Materials, ETH Z{\"u}rich, Vladimir-Prelog-Weg 5, 8093 Z{\"u}rich, Switzerland}
	
	\author{Fabio Grillo}
	\affiliation{Laboratory for Soft Materials and Interfaces, Department of Materials, ETH Z{\"u}rich, Vladimir-Prelog-Weg 5, 8093 Z{\"u}rich, Switzerland}
	
	\author{Gerhard Gompper}
	\affiliation{Theoretical Physics of Living Matter, Institute of Biological Information Processing and Institute for Advanced Simulation, Forschungszentrum J{\"u}lich, 52425 J{\"u}lich, Germany}
	
	\author{Lucio Isa}
        \email{E-mail: lucio.isa@mat.ethz.ch}
	\affiliation{Laboratory for Soft Materials and Interfaces, Department of Materials, ETH Z{\"u}rich, Vladimir-Prelog-Weg 5, 8093 Z{\"u}rich, Switzerland}

	\date{\today}

	\begin{abstract} 
	  The generation of fluid flows by autophoretic microswimmers has been proposed as a mechanism to enhance mass transport and mixing at the micro- and nanoscale. Here, we experimentally investigate the ability of model 2-D “active baths" of photocatalytic silica-titania Janus microspheres to enhance the diffusivity of tracer particles at different microswimmer densities below the onset of collective behaviour. Inspired by the similarities between our experimental findings and previous results for biological microorganisms, we then model our Janus microswimmers using a general squirmer framework, specifically treating them as neutral squirmers. The numerical simulations faithfully capture our observations, offer an insight into the microscopic mechanism underpinning tracer transport, and allow us to expand the parameter space beyond our experimental system. We find strong evidence that near-field interactions dominate enhancements in tracer diffusivity in active Janus baths, leading to the identification of an operating window for enhanced tracer transport by chemical microswimmers based on scaling arguments. Based on this argumentation, we suggest that for many chemically active colloidal systems, hydrodynamics alone is likely to be insufficient to induce appreciable mixing of passive components with large diffusion coefficients.
	\end{abstract}
	
	\pacs{???}
	
	\maketitle


\section{Introduction}
The ability of active matter to transduce available energy into directed motion has inspired research across various disciplines \cite{Ramaswamy2010,Gompper2020}. Amongst the systems studied at the micron-scale, where Brownian fluctuations and viscous forces are significant \cite{Bechinger2016}, the “autophoretic" motion of chemical microswimmers has been proposed to induce micro-mixing in fluids \cite{Karshalev2018,Dai2022}, with potential applications ranging from environmental remediation \cite{Wang2019} to synthetic chemistry \cite{Wittmann2022}. Enhancing diffusivity at the micro-scale has been extensively researched in the context of biological microswimmers \cite{Wu2000,Leptos2009,Valeriani2011,Wilson2011,Lin2011,Morozov2014,Jeanneret2016,Lagarde2020}, with possible implications including enhanced nutrient uptake by the micro-organisms \cite{Kanazawa2020} and even ocean biomixing \cite{Thiffeault2010}. At even smaller length scales, “active enzyme solutions" have been shown to enhance tracer diffusion via momentum transfer \cite{Mikhailov2015,Zhao2017}. Nevertheless, analogous systematic studies of enhanced mixing by synthetic, catalytic microswimmers are lacking \cite{Mino2011}, except for systems propelled by bubble generation \cite{Wang2019c,Orozco2014}. The mixing capabilities of such systems are nevertheless of interest, e.g. for advanced reactor designs incorporating microswimmers, where the use of biological materials or the formation of bubbles may be undesirable, but where mass-transport limitations are important.

Here, we investigate the mixing properties of a widely studied synthetic active matter model system consisting of a suspension of photocatalytically active Janus particles with a titania cap under UV illumination. We overcome the limited scale of conventional Janus particle fabrication techniques via `Toposelective Nanoparticle Attachment” (TNA), which exploits an emulsion template and multi-functional polymers to facilitate the selective binding of a broad range of nanoparticles, including the photocatalytic titania nanoparticles used in this study, onto silica particle supports \cite{Bailey2021a,bailey2022_mod}. The advantages of TNA not only lie in the greater than 100 mg yields per batch, which ensures reproducibility between experiments \cite{Wittmann2021}, but also the scalability and modularity offered in catalyst selection, which is desirable from an applications perspective. While experimentally probing the mechanisms leading to the enhanced transport of passive tracers, we observe parallels between their dynamics in active baths of our synthetic active colloids and previous findings for biological microswimmers \cite{Leptos2009,Thiffeault2015,Jeanneret2016} - despite their vastly different underlying propulsion schemes. Specifically, the passive particles demonstrate Lévy flight-like trajectories \cite{Kanazawa2020}, characterised by diffusive displacements inter-dispersed with large translational “jumps". 

Drawing on the concept of “active flux" \cite{Mino2011,Mino2013,Jepson2013}, we determine a positive correlation between the time-scale expected for swimmers to interact with the passive tracers via hard-core and near-field hydrodynamic interactions \cite{Valeriani2011,Lin2011,Pushkin2013,Thiffeault2015,Burkholder2017}, and the peak in the non-Gaussian parameter attributable to the tails in the distribution of displacements arising from tracer jumps \cite{Jeanneret2016}. We then use a coarse-grained hydrodynamic swimmer model - a “squirmer"  which is equivalent to a force dipole and source dipole in the far-field - to capture the generic aspects of tracer mixing \cite{Gotze2010,Zottl2023}. Noting the good qualitative agreement between numerical simulations and experiments, we proceed to study the mixing properties of our hydrodynamic squirmers under conditions beyond those experimentally achievable, and observe a linear scaling in the diffusivity enhancement of our tracers with the active flux of the squirmers \cite{Wu2000,Leptos2009,Mino2013}. Making use of dimensionless quantities, we conclude by highlighting several key considerations for the design of catalytic "micro-stirrers".

\section{Results}
\subsection{Experimental system}
\begin{figure}
\centering
  \includegraphics[width=\columnwidth]{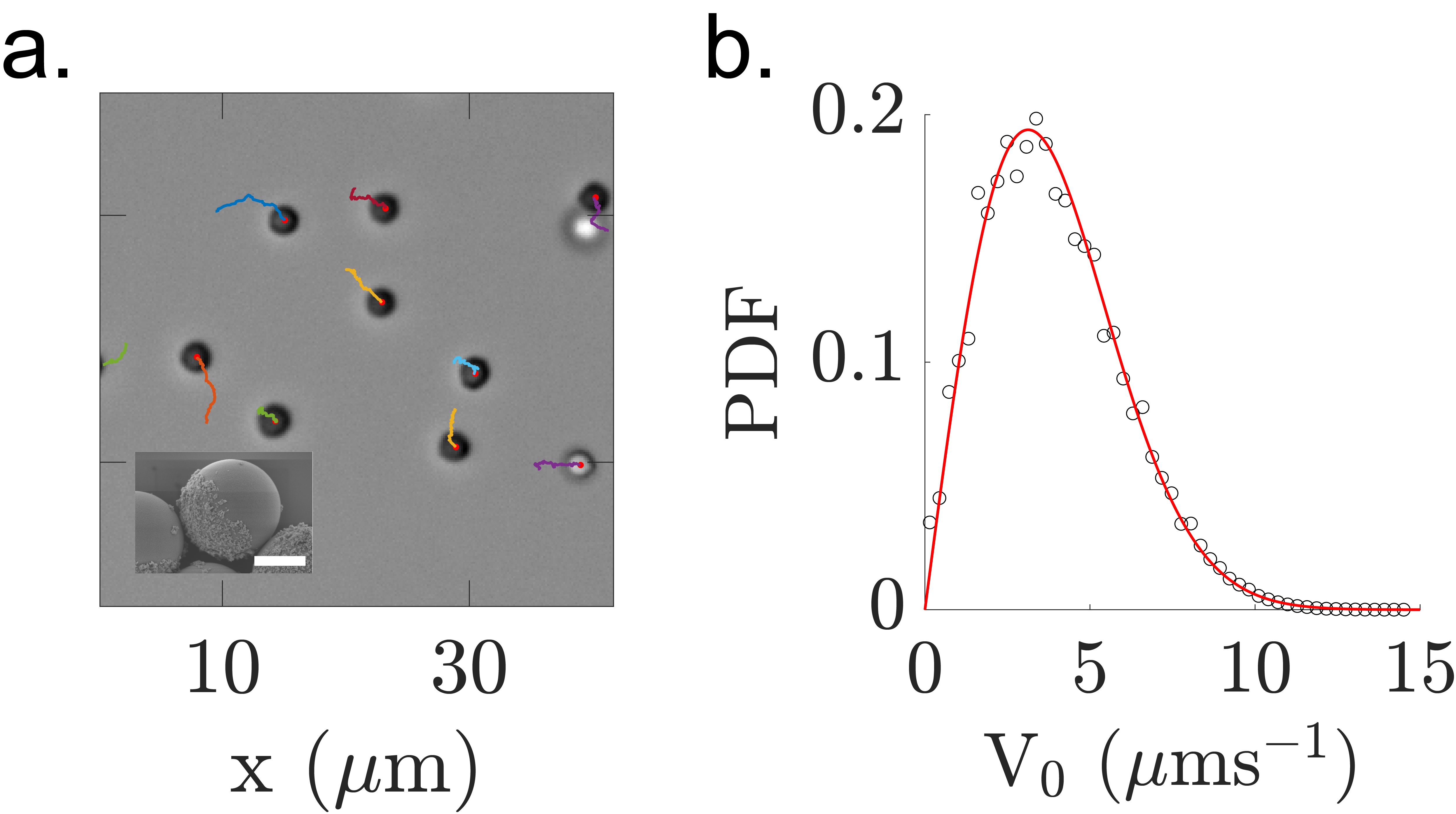}
  \caption{Janus microswimmer 2D motion above a substrate in the absence of clustering. a. Micrograph of “free" particles at low area fraction $\Phi_{swim}$ = 0.02 (1 s trajectory). Inset: HR-SEM image of the studied Janus particles. Scale bar represents 1 $\mu$m.  b. Distribution of the free microswimmer in-plane speeds measured over 40000 trajectories.}
  \label{fig:Fig1}
\end{figure}

Our microswimmers are spherical SiO\textsubscript{2} colloids (diameter $d_{swim} = 2.16 \mu$m) half-decorated with photocatalytic TiO\textsubscript{2} nanoparticles, as described in detail in Refs. \cite{Bailey2021a,bailey2022_mod}. Under UV illumination, the particles asymmetrically degrade H\textsubscript{2}O\textsubscript{2} (fuel) and thus establish local chemical gradients, leading to self-propulsion by autophoresis \cite{Howse2007,Golestanian2009,Popescu2010,Dey2016,Bailey2021b}. The density mismatch of the particles, coupled with the experimental conditions used (see Supporting Information, Materials \& Methods) minimises their out-of-plane motion away from the substrate \cite{Bailey2022_ML,Bailey2024} and ensures that particles move in two dimensions (2D xy plane). Under these conditions, we systematically study the behavior of the microswimmers as their effective area fraction $\Phi_{swim} = N_{swim}A_p/A_T$, where $N_{swim}$ is the number of microswimmers, $A_p = \pi d_{swim}^2/4$ is the 2-D projection of the microswimmer area, and $A_T$ is the area of the system, is varied. In agreement with previous reports for both synthetic and biological systems \cite{Theurkauff2012,Thakur2012,Palacci2013,mognetti_2013,Liebchen2017,Ginot2018,chen_2012,chen_2015,Theers2018}, we observe the formation of ``dynamic clusters” above a threshold swimmer density (see Supporting Information, Figure S1, and Videos S1 and S2). 
The emergence of collective behaviour, although of significant fundamental interest, not only complicates the analysis of the mixing properties of microswimmers, but also likely hinders their overall mixing efficiency \cite{Mou2016}, as we outline below. We therefore determine the limiting area fraction of microswimmers before the onset of clustering events using the methodology outlined by Theurkauff et al. \cite{Theurkauff2012} (see Supporting Information, Figure S1a). In agreement with their findings, we determine an onset of clustering at $\Phi_{swim} \approx 0.05$ (see Supporting Information, Figure S1c), which we thus set as the upper bound for our experiments to focus on single-particle effects. Similar to the seminal work by Wu and Libchaber \cite{Wu2000} on the mixing properties of bacterial baths, the speed distribution of our synthetic ``free” particles obeys Maxwell’s statistics,  however, in its 2D form as particle motion is confined above the substrate: $P(V_0) = \frac{V_0}{v_M^2}\text{exp}(-\frac{V_0^2}{2v_M^2})$, where $v_M$ is the modal speed. Under the experimental conditions used, our active colloids have a modal speed $v_M \approx 3.1\mu$ms\textsuperscript{-1} (see Figure \ref{fig:Fig1}b). 

\subsection{Tracer dynamics}
Having established the upper bound for the number density of our microswimmer baths, we investigate the effect of microswimmers on the dynamics of micron-sized tracers (diameter $d_T = 1.70 \mu$m). Notably, we find that the tracers display extensive displacements when a microswimmer passes in their vicinity (Figure 2a, Supporting Information Video S3) - which we henceforth refer to as “jumps". We underline that these enhancements in the motion of the tracers are distinct to previous studies on colloidal transport, where the microswimmer actively transports the tracer by making use of phoretic or hydrodynamic effects to drag the tracer along its trajectory \cite{Baraban2012,Palacci2013b,mou_2019} (also see Supporting Information Figure S2 on the over-weighted effect such transported tracers can have on the ensemble mean-squared-displacement (EMSD)). In contrast, the jumps do not require that the tracer is effectively carried by the microswimmer, and instead extended displacements are also observed in directions distinct from the velocity of the microswimmers (see Supporting Information, Video S3).The presence of these large displacements, which result in trajectories reminiscent of Levy flights \cite{Kanazawa2020}, is well established for passive tracers in the presence of different suspensions of micro-organisms \cite{Wu2000,Leptos2009,Jepson2013,Jeanneret2016}, and is attributed to the velocity flow fields created by their motion \cite{Thiffeault2010,Lin2011,Morozov2014}. We however note the absence of previously observed loops in our observed tracer trajectories, potentially arising from the flow-distorting effect of the wall above which the particles move \cite{Mino2013}, or from the effect of near-field hydrodynamic and steric interactions \cite{Valeriani2011,Pushkin2013,Jeanneret2016,Mathijssen2018,Lagarde2020}. 

\begin{figure*}
\centering
  \includegraphics[width=\linewidth]{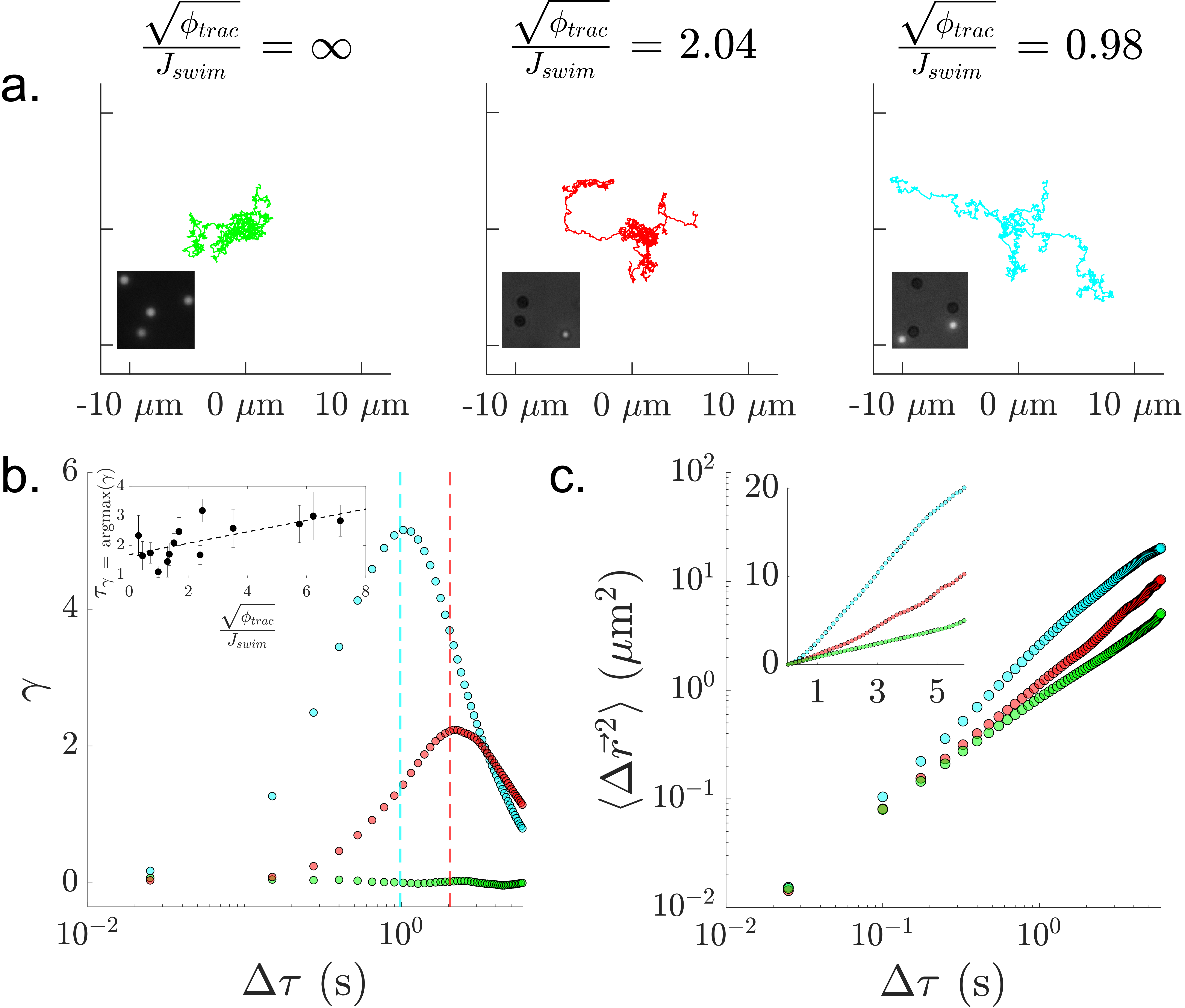}
  \caption{Emergence of large displacements in the trajectories of tracers with the introduction of microswimmers. a. Tracer trajectories for 3 different experimental cases corresponding to various $\sqrt{\phi_{trac}}/J_{swim}$ (green: infinity, red: 2.04 s, cyan: 0.98 s). Scale bars represent 5 $\mu$m. Insets are representative snapshots of the experimental system (bright tracers and dark microswimmers, 16.25 x 16.25 $\mu$m). b. Evolution of the kurtosis $\gamma$ of the distribution of displacements with lag time $\Delta\tau$ for the 3 experimental conditions depicted in a. Each data-set contains at least 23 trajectories. The dashed lines indicate $\sqrt{\phi_{trac}}/J_{swim}$ for the given experimental conditions. Inset: Correlation between the estimated near-field interaction time $\sqrt{\phi_{trac}}/J_{swim}$ between a tracer and a swimmer and the experimentally observed lag time at which the kurtosis is maximised. Error bars indicate the standard error of the mean. c. Ensemble mean-squared-displacement (EMSD) of the 3 experimental systems depicted in a.}
  \label{fig:Fig2}
\end{figure*}
\subsection{Hydrodynamic mixing via active flux}

Via simple scaling arguments, we now seek to demonstrate that these tracer jumps are at least in-part caused by near-field hydrodynamic and hard-core interactions between the tracers and the microswimmers.
We begin with the concept of the “active flux" of microswimmers, proposed by Mino and co-workers \cite{Mino2011,Mino2013} as $J_{swim} = \phi_{swim}\cdot\langle V_0\rangle$, where $\phi_{swim} = N_{swim}/A_T$ is the number density of microswimmers and $\langle V_0\rangle$ is their mean speed. Assuming that the $N_{trac}$ passive tracers in the system are equidistantly distributed, it follows that the characteristic length-scale between tracers can be written as $1/\sqrt{\phi_{trac}}$, where $\phi_{trac}$ is the number density of tracer particles. Given these definitions, we then identify a characteristic time $\tau_c = \sqrt{\phi_{trac}}/J_{swim}$ for near-field interactions between a microswimmer-tracer pair, i.e. the time expected for a swimmer to travel the inter-tracer distance given the activity flux. Returning to the tracer jumps, these larger displacements are expected to dominate the higher orders of the distribution of tracer displacements, i.e. its kurtosis ($\gamma = \langle(\Delta x - \langle\Delta x\rangle)^4\rangle/\langle(\Delta x - \langle\Delta x\rangle)^2\rangle^2 - 3$), due to their higher weighting for larger powers \cite{Thiffeault2015,Jeanneret2016}. Here, $\Delta x$ denotes the one dimensional (1D) independent displacements of the tracers in x and y, which are then binned together for increased statistics. Extracting $\gamma$ from the tracers for different experimental realisations of $\tau_c$, we find a positive correlation between $\tau_c$ and $\tau_{\gamma=\operatorname*{argmax}_\tau(\gamma)}$, the lag-time $\tau$ at which $\gamma$ is at a maximum (see Figure \ref{fig:Fig2}b). As $\tau_c$ decreases, and swimmer-tracer “collisions" become more prevalent, the overall diffusivity of the tracers is also enhanced, as can be detected in their ensemble mean-squared-displacement (EMSD) (see Figure \ref{fig:Fig2}c). 

\begin{figure*}
\centering
  \includegraphics[width=\linewidth]{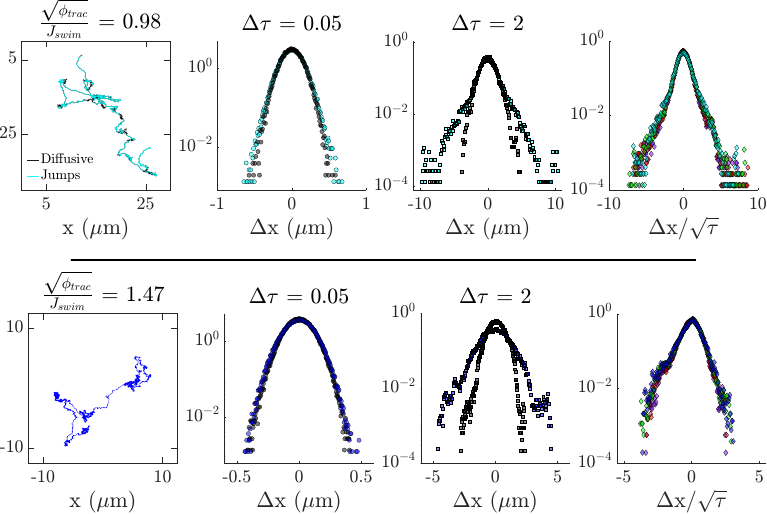}
  \caption{Influence of the jumps (cyan and dark blue for experiments and simulations respectively) on the tracer distribution of displacements at different lag times $\Delta\tau$. Top row: Experimental tracer trajectory in a microswimmer bath. The non-Gaussian, exponential tails in the distribution of displacements becomes prominent at longer lag-times. We note that these tails are not observed for “passive" microswimmers (see Supporting Information, Figure S3). Bottom row: Our numerical neutral squirmer model maps qualitatively well onto the experimental findings, with the presence of jumps at longer $\Delta\tau$ seen in the tails of the distribution of displacements. In both experiments and simulations, we observe “diffusive scaling" with $\sqrt{\tau}$ \cite{Leptos2009}, which is attributed to the intermediate values of $\tau$ studied where the near-field dominates \cite{Thiffeault2015}. Violet, red, and green diamonds correspond to $\tau = 0.75, 1, 1.5$s respectively ($\tau = 2$s indicated by cyan and blue diamonds for experiment and simulation as before).
  }
  \label{fig:Fig3}
\end{figure*}

Further evidence for the enhancement in tracer diffusivity as a result of tracer jumps is presented in Figure \ref{fig:Fig3} (top row). Following Jeanneret et al. \cite{Jeanneret2016}, we distinguish between standard diffusive motion during a fixed time and jumps, using the autocorrelation of subsequent displacements. Although the significantly lower speeds of the studied synthetic microswimmers reduces the effective size of the tracer jumps, it is clear that long-time exponential tails emerge in the distribution of tracer displacements when jumps are included, replicating the findings for microalgae \cite{Jeanneret2016}. Re-scaling the distribution of displacements at intermediate lag-times by the square root of the lag-time $\sqrt{\tau}$ causes their collapse onto a single curve. This “diffusive scaling", first observed in active baths of eukaryotic micro-organisms \cite{Leptos2009}, is attributed to the effect of near-field interactions which cause large displacements, i.e. jumps, which dominate higher order moments of the distribution at intermediate times \cite{Thiffeault2015}. We therefore hypothesise that near-field interactions between Janus microswimmers and passive particles, resulting in large tracer displacements, play a dominant role in enhancing their diffusivity. Far-field flows, if generated by the microswimmers, may in fact be dampened by the presence of the underlying substrate, which can dissipate the generated flows \cite{Mino2013,Morozov2014}. 

The qualitative agreement between our findings and those for micro-organisms was not expected \textit{a-priori}, but provides us with the opportunity to use theory developed for such systems to describe our own. Specifically, the squirmer model \cite{Gotze2010}, first proposed for microorganisms \cite{Lighthill1952,Blake1971}, is frequently invoked to describe the flow fields surrounding chemical Janus microswimmers \cite{Michelin2014,Zottl2023,Bailey2024}, and generates flows which are consistent with a coarse-grained framework incorporating hydrodynamic effects in the far field via a multipole expansion, together with appropriate near-field fluid flows. We note that a more complete description of chemically active colloids requires the analysis of the phoretic flows resulting from the chemical gradients produced by the microswimmers \cite{Popescu2018a,Katuri2021} (see Supporting Information, Figure S4). Still, a complete description of the phoretic and osmotic flows requires the relevant mobility parameters - reducing generalisability to other mixing systems. For a coarse-grained model to describe the mixing properties of our microswimmers, we therefore neglect the phoretic contribution to tracer displacements, following Ref. \cite{Campbell2019} (also see Supporting Information, Figure S5). 

\subsection{Numerical simulations}

We begin by modelling our synthetic microswimmers as neutral squirmers (squirmer parameter $\beta$ = 0 - see Supporting Information, Materials and Methods for a detailed description of the simulation method and description of parameters), informed by previous work on catalytic Janus particles \cite{Yang2014}, and the dominant role of near-field interactions predicted for such systems in enhancing tracer diffusivity \cite{Lin2011,Pushkin2013}. The neutral squirmer model furthermore agrees with the absence of transport behaviours associated with flow vortices \cite{Madden2022} and the absence of long-range forces observed \cite{Pushkin2013} (see Supporting Information, Figure S5). Fluid flow is modelled using dissipative particle dynamics (DPD), a mesoscopic hydrodynamics simulation technique where the fluid is represented by a large but finite number of solvent particles. In DPD, stochastic and dissipative forces are introduced to act as a thermostat \cite{Hoogerbrugge1992,Espanol1995}. Here, both passive and active colloids consist of particles placed homogeneously on a spherical surface, connected by bands to form a triangulated mesh, to ensure a near-rigid shape (see Supporting Information, Materials and Methods for more details). Propulsion is modelled by imposing a slip velocity at the surface of the active particles using the spherical squirmer model \cite{Alarcon_MCP_2017}, and solving Newton's equations of motion using the velocity-Verlet algorithm. For passive particles, no-slip boundary conditions (BCs) are imposed. To relate simulation and physical units, we introduce characteristic length and time scales. The length scale is based on the squirmer diameter $d_{Sq} = 2.16 \mu$m, as in experiments, and the time-scale is set by $\tau_{sq} = d_{Sq}/v_M$, where $v_M$ is the modal speed of the speed distribution of active particles in the xy plane previously calculated (see Figure \ref{fig:Fig1}b), which is set as the squirmer speed. In this manner, the simulation domain is set to 33.2 x 33.2 x 6.8 $\mu$m with periodic BCs in the xy plane, and walls at z = 0 and z = 6.8 $\mu$m. Rigid walls are simulated by immobilised DPD particles having identical number density and interactions  as the fluid particles. The immobile wall particles enforce no-slip boundary conditions (BCs) at the walls through dissipative interactions with the fluid particles. In order to prevent wall penetration by fluid particles, reflective surfaces are introduced at the interface between the fluid and the walls. Both passive and active particles are confined within a layer of 2.45 $\mu$m at the lower wall with no-slip BCs, to mimic the quasi 2D setup in the experiments. The wall at z = 6.8 $\mu$m assumes slip BCs, to better capture experimental conditions. The area fraction of active particles corresponds to $\Phi_{swim}$ = 0.0597, identified as the approximate onset of clustering in both experiments and simulations.

\begin{figure}
\centering
  \includegraphics[width=\columnwidth]{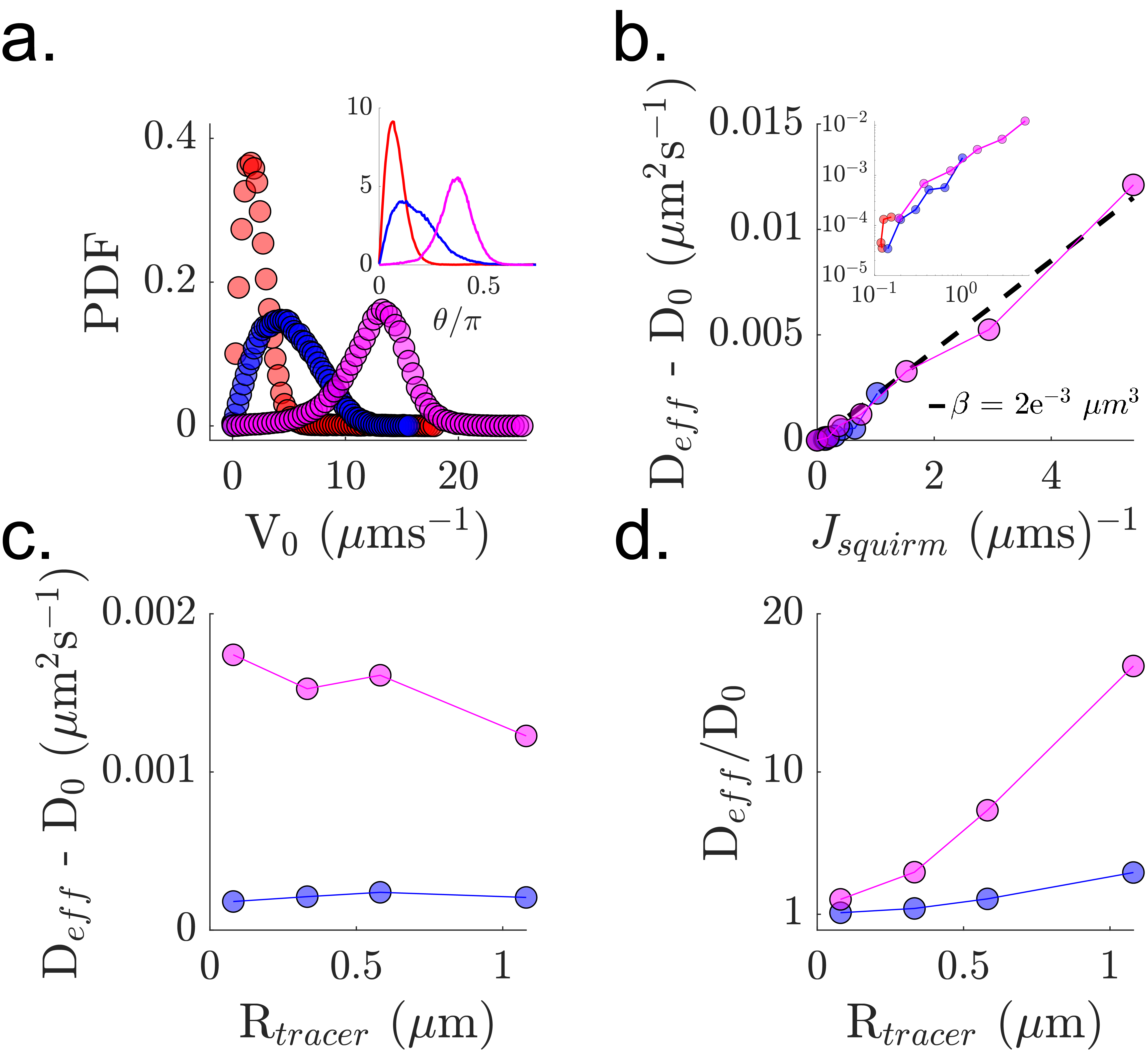}
  \caption{Dynamics of simulated active and tracer particles (red for pullers, blue for neutrals, and magenta for pushers.) a. Distributions of squirmer in-plane (2D) speeds for fixed squirmer strengths (all types would have the same swimming speed under unbounded conditions). Inset: Orientation distributions of active and tracer particles with respect to the z axis (i.e. perpendicular to the wall). b. Increase in the diffusion coefficient of the tracers, D$_{eff}$ - D$_0$, with increasing squirmer flux $J_{swim}$. The dashed line indicates a linear least-squares fit for the pushers. Inset: log-log plot for visualisation of data points. c. The enhancement in the tracer diffusion coefficient, D$_{eff}$ - D$_0$, is relatively flat across different tracer sizes \cite{Leptos2009,Mino2011,Mino2013,Morozov2014}, for pushers, the slight increase with decreasing size may arise from the growing role of tracer entrainment \cite{Pushkin2013,Morozov2014}. d. Normalised enhancement in the diffusion coefficient, D$_{eff}$/D$_0$, for different sizes of passive tracers.}
  \label{fig:Fig4}
\end{figure}

We find that the dynamics of our simulated passive tracer particles in the presence of neutral squirmers are qualitatively very similar to our experimental findings under the same conditions. Specifically, we identify the presence of jumps in the tracer trajectories, which emerge at longer lag times in the distribution of tracer displacements, as well as their diffusive scaling at the same intermediate lag-times investigated experimentally (see Figure \ref{fig:Fig3}, bottom row). We note that the magnitude of the displacements determined in numerical simulations are approximately half that observed in experiment, which may be attributed to the role of phoretic flows or finer details of the near-field hydrodynamics of our synthetic microswimmers not incorporated in our coarse-grained model. Nevertheless, encouraged by the good agreement between experiment and simulation, we proceed to a more quantitative characterisation of the mixing efficiencies of different squirmers, not only for neutral squirmers ($\beta$ = 0), but now also for pullers ($\beta$ = 5), and pushers ($\beta$ = -5). Here, $\beta = B_2/B_1$, where $B_2$ introduces a fore-aft asymmetry in the slip velocity field of the squirmer, and $B_1$ governs the propulsion strength of the microswimmer (see Supporting Information, Materials and Methods for more details). As a control, we also study the dynamics of tracer particles in the presence of particles that do not swim (i.e. passive particles). For a given propulsion strength $B_1$ selected to mirror experimental speeds, we find that pushers are superior mixers to other squirmer types under the same numerical conditions. However, we note that for a given propulsion strength, the three squirmer types can have significantly different swimming speeds (see Figure \ref{fig:Fig4}a). Upon closer inspection, we find that the discrepancy in swimming speeds is a result of their orientation with respect to the substrate (see Figure \ref{fig:Fig4}a, inset). It is well known that pushers will align parallel to a substrate (resulting in higher speeds), whereas pullers will orient perpendicular to surfaces \cite{Spagnolie2012}. Neutral squirmers in contrast possess characteristics of both squirmer types, reflected in their intermediate orientation with the substrate, resulting in a distribution of swimming speeds mirroring that observed for our experimental synthetic microswimmers (compare Figures \ref{fig:Fig1}b and \ref{fig:Fig4}a). By instead comparing the long-term enhanced diffusion coefficient D$_{eff}$ - D$_0$ (where D$_{eff}$ is the coefficient in the presence of the swimming squirmers, and D$_0$ is the diffusion coefficient measured in the absence of squirmer activity) with the active flux of the squirmer $J_{squirm}$, we observe a collapse onto the expected linear scaling given by D$_{eff}$ - D$_0$ = $\alpha J_{squirm}$ \cite{Mino2011,Mino2013,Jepson2013,Kanazawa2020}, with $\alpha \approx 2e^{-3}$ m\textsuperscript{3} or (0.126 m)\textsuperscript{3} (see Figure \ref{fig:Fig4}b). We note that this value for $\alpha$ is approximately one order of magnitude less than that previously calculated for active rods and bacteria \cite{Mino2011,Mino2013}.

We then study the effect of reducing the tracer size, $R_{tracer}$, for squirmers with swimming speeds approximately equal to that determined for our Janus microswimmer system. In agreement with previous investigations \cite{Mino2011,Mino2013,Jepson2013,Morozov2014}, we find that tracer size has a relatively small effect on the increase in its long-term diffusion coefficient for a given activity flux (see Figure \ref{fig:Fig4}c). The small enhancement observed for decreasing tracer sizes in the case of pushers could be attributed to the mechanism of entrainment, as the smaller tracers are able to closer approach the squirmer body \cite{Pushkin2013,Morozov2014}. However, upon re-scaling the enhanced tracer diffusivity D$_{eff}$ by the underlying diffusivity of the tracers D$_0$, we see a clear reduction in the ability of the squirmers to “mix" the tracers (see Figure \ref{fig:Fig4}d). We thus hypothesise that the mixing capacity of microswimmers is limited by the relative size and frequency of the imparted “jumps" via near-field interactions compared to the underlying diffusivity of the component to be mixed (i.e. $l_j \gg \sqrt{\text{D}_0\tau_c}$ for enhanced mixing, where $l_j$ is the jump length after an interaction occurring with frequency $1/\tau_c$). This mirrors the experimental findings using 230 nm nanoparticle tracers where no enhancement in $D_{eff}$ over $D_0$ was observed (see Supporting Information Figures S6-7), suggesting the unsuitability of this type of microswimmer systems to mix e.g. species with very high diffusion coefficients.

\subsection{Mass-transport considerations}
Additional challenges facing the application of micron-scale synthetic microswimmers to overcome mass-transport limitations can be identified by following Purcell's example and evaluating the relevant dimensionless numbers \cite{Purcell1977,Berg1977,Purcell1978}. Presuming low Reynolds number conditions, i.e. $Re = vl/\nu \ll 1$, where $\nu$ is the kinematic viscosity and $vl$ relates to the fluid flow at a characteristic length-scale $l$, for microswimmers swimming with speed $v$ (i.e. motion in the viscous regime and ignoring substrate effects), we begin with the strength of swimming to thermal diffusion, $Pe_T = vl/D$, where $D$ is the translational diffusion coefficient of the transported species. Ignoring hydrodynamic interactions (i.e. only accounting for activity and excluded volume effects), Burkholder and Brady \cite{Burkholder2017} demonstrated that for large values of $Pe_T$, steric “collisions" between the active Brownian particles and passive probe lead to enhanced diffusion. However, if $Pe_T \ll 1$ - as in many proposed applications for “chemistry-on-the-fly" \cite{Karshalev2018,Dai2022,Wittmann2022} - the microswimmer is essentially static with respect to the transported species, and we do not expect any diffusivity enhancement. In fact, the “swimming catalysts" may be better off to remain stationary and wait for the diffusion of the reagents \cite{Kanazawa2020}, a point already well elaborated upon by Purcell and Berg \cite{Purcell1977,Berg1977,Purcell1978}. 

Nevertheless, allowing reasonable values of $Pe_T$, we then evaluate the Damköhler $Da$ and Sherwood $Sh$ numbers. $Da = \kappa l/D$ details the balance between the reaction rate $\kappa$ and the (diffusive) mass transfer process. Notably, the classical continuum phoretic framework to describe chemical microswimmer motion presumes reaction-rate limited kinetics, i.e. $Da\rightarrow 0$ \cite{Howse2007,Kreissl2016,Popescu2018a,Dominguez2022}, with finite values of $Da$ leading to decreasing microswimmer speeds \cite{Michelin2014} - calling into question the goal of overcoming mass-transfer limitations in reactions using microswimmer motility \cite{Karshalev2018,Dai2022,Wittmann2022}. Regardless, assuming transport-limited reactions, we examine the Sherwood number for mass transfer to a sphere in Stokes flow, $Sh = 2 + 0.991Re^{1/3}Sc^{1/3}$ \cite{Friedlander1961,Armenante1989}, where $Sc = \nu/D$ is the Schmidt number (the ratio of momentum to mass diffusivity), the first term is for the asymptotic limit of molecular diffusivity, and $0.991Re^{1/3}Sc^{1/3}$ is the advective contribution to mass-transport \cite{Armenante1989}. In creeping flow, we thus expect the enhancement in mass transfer from fluid advection to be small even when $Pe_T>1$ for most physical parameters. In highly viscous reactor settings or when mixing macromolecules with a low diffusivity (or situations combining both features \cite{RuizGonzlez2024}), the micro-stirrer strategy could be more effective, especially where the microswimmers are large \cite{Berg1977,Purcell1978}. For example, Kim and Breuer \cite{Kim2004} demonstrated that \textit{E. Coli} could enhance the diffusivity of 77 kDa Dextran ($\approx 10$nm spheres assuming spheres with Stokes-Einstein diffusion) by a factor 4 in a microfluidic chip. Nevertheless, we underline the poor scaling of the Schmidt number with the exponent $1/3$, and question whether such modest enhancements in mixing would justify complex reactor designs incorporating motile catalysts. 

\section{Discussion and conclusions}

Basic dimensionless arguments based on convection-reaction-diffusion theory \cite{Purcell1977,Berg1977,Purcell1978,Squires2008} thus provide some additional insight into the challenges of using autophoretic microswimmers as micro-stirrers, besides the limitations posed by neglected emergent phenomena such as clustering. Still, it is worth noting that collective behaviours could also enhance fluid intermixing and thereby improve reaction-diffusion limitations \cite{Wang2021a}, while appropriate material selection can modify generated flow fields \cite{Katuri2018a,Sharan2022} and thereby prevent undesirable aggregation (i.e. for pushers \cite{Theers2018,Qi2022}). Other promising avenues to increase microswimmer mixing could be achieved via modifying their shape away from classical spheres \cite{xiong_2023}, exploiting bubble generation \cite{MundacaUribe2023}, or multi-step procedures making use of different populations of microswimmers \cite{RuizGonzlez2024}. Such strategies could in particular be of use for “high-value" applications where lower quantities of micro- or nanoswimmers are required and cost is of lesser concern, e.g. for medical purposes. Our results therefore underline the need for a more considered approach to the design and proposed application of synthetic microswimmers to enhance micro- and nanoscale mass transport, and provide evidence that deviations from model (spherical) swimmers may be required for such purposes.

\medskip
\textbf{Acknowledgements} \par
The authors thank M. Popescu and V. Niggel for helpful discussions on theory and experiment respectively. The authors also thank S. Vasudevan for his assistance with the performed DDM analysis. The authors gratefully acknowledge computing time on the supercomputer JURECA \cite{Thornig2021} at Forschungszentrum J{\"u}lich under grant no. actsys.  

\medskip
\textbf{Author Contribution Statement} \par
Author contributions are defined based on the CRediT (Contributor Roles Taxonomy). Conceptualisation: M.R.B., D.A.F., F.P., F.G., G.G., L.I. Funding acquisition: L.I. Investigation and analysis: M.R.B., D.A.F. Methodology: M.R.B., D.A.F., F.P. Supervision: L.I. Visualisation and writing: M.R.B., D.A.F., F.P., F.G., G.G., L.I.

\bibliographystyle{apsrev4-1}
\bibliography{Paper4_Micromixing}

\onecolumngrid

\newpage

\renewcommand{\figurename}{Figure. S}
\renewcommand{\tablename}{Table S}

\setcounter{figure}{0}
\setcounter{table}{0}

\section*{The Supporting Information includes:}

\begin{itemize}[nosep]
    \item[] Text S1: Materials and Methods
    \item[] Fig. S1: Definition of clustering and the different speed distributions
    \item[] Fig. S2: Influence of actively transported tracers on ensemble averaged values
    \item[] Fig. S3: Distribution of tracer displacements in the presence of non-swimming microswimmers (UV light off)
    \item[] Figure S4: Evidence for phoretic interactions between the Janus particles (pumps) and passive nanoparticle tracers
    \item[] Figure S5: Influence of Janus pumps on nanoparticle tracer dynamics in the near- and far-field
    \item[] Figure S6: Lack of mixing effect observed for nanoparticle tracers in a microswimmer bath
    \item[] Figure S7: Comparison of different approaches to evaluate nanoparticle tracer dynamics to evaluate their accuracies
    \item[] Video S1: Motion of microswimmers at low number density, below the threshold for dynamic clustering. The playback speed is 2x real time, and the length of acquisition is 5s.
    \item[] Video S2: Motion of microswimmers at high number density, above the threshold for dynamic clustering. The playback speed is 2x real time, and the length of acquisition is 5s.
    \item[] Video S3: Dynamics of passive tracers (fluorescent, white) in the presence of microswimmers (non-fluorescent, dark). We note that near-field interactions appear to dominate the enhancement in passive tracer displacement. The playback speed is 1x real time, and the length of acquisition is 7.9s.
 
\end{itemize}

\newpage

{\huge Materials and Methods}\\

\noindent
{\bf Experimental setup and materials.} Particles were imaged on an inverted microscope (Nikon Eclipse Ti2e) under Köhler illumination with white light using a 40× objective (CFI S Plan Fluor ELWD 40XC) with adjustable collar (set to 1 mm) unless otherwise specified. Due to a density mismatch, the particles sediment to the glass substrate, where they remain under the experimental conditions used. Images were acquired at 50 FPS using a Hamamatsu C14440-20UP digital camera. The Janus microswimmers were activated with a UV LED (365 nm, Prizmatix UHP-F-365), at a power density of 44 mW/mm\textsuperscript{2}, in the presence of a chemical fuel (H\textsubscript{2}O\textsubscript{2} 3 \% v/v, Acros Organics). The tracer particles (230 nm and 1700 nm diameter, microParticle GmbH) were fluorescent under illumination with the UV LED source, and did not require an additional excitation source (Ex/Em 502/518 nm). To ensure a near-constant number density (area fraction $\Phi$) of studied microswimmers, it was necessary to create a confined system to prevent their escape to non-illuminated regimes. This was achieved by creating wells (diameter = 250 $\mu$m) in glass slides by HF etching. To avoid boundary effects, only the region bounded by a box (83.2 x 83.2 $\mu$m) centred in this well was studied. The cells were sealed using an upper glass slide and PDMS spacer to prevent convective drift.

\noindent
{\bf Modeling fluid flow in simulations.} Fluid flow is modeled by the dissipative particle dynamics (DPD) method \cite{Hoogerbrugge1992,Espanol1995},
which is a mesoscopic hydrodynamics simulation technique. In DPD, fluid is represented by a collection of $N_s$
solvent particles. DPD particle interactions are pairwise and soft, while particle
speeds ${\bf v}_i$ and positions ${\bf r}_i$ follow the Newton's second law
of motion
\begin{eqnarray}\nonumber
\frac{d {\bf r}_i}{dt}  & =  & {\bf v}_i, \\
\frac{d {\bf v}_i}{dt} & = & \frac{1}{m_i}\sum_{j \neq i} ({\bf F}^C_{ij} + {\bf
F}^D_{ij} + {\bf F}^R_{ij}),
\end{eqnarray}
where $m_i$ is the particle mass and the forces ${\bf F}^C_{ij}$, ${\bf F}^D_{ij}$,
and ${\bf F}^R_{ij}$ define conservative, dissipative, and random interactions,
respectively. The above equations of motion are integrated using the velocity-Verlet algorithm.
The total force exerted on a particle $i$ by particle $j$ is given by
\begin{eqnarray}\nonumber
{\bf F}^C_{ij} & = & b\left(1-\frac{r_{ij}}{r_c} \right) \hat{\bf r}_{ij},  \\  \nonumber
{\bf F}^D_{ij} & = & -\gamma \omega^D(r_{ij})({\bf v}_{ij} \cdot \hat{\bf r}_{ij}) \hat{\bf r}_{ij}, \\
{\bf F}^R_{ij} & = & \sigma_{dpd} \omega^R(r_{ij}) \frac{\xi_{ij}}{\sqrt{dt}} \hat{\bf r}_{ij},
\label{eq:dpd_forces}
\end{eqnarray}
where ${\bf r}_{ij}={\bf r}_i-{\bf r}_j$, $\hat{\bf r}_{ij}={\bf r}_{ij}/r_{ij}$,
and ${\bf v}_{ij}={\bf v}_i-{\bf v}_j$. The coefficients $b$, $\gamma$, and $\sigma_{dpd}$
determine the strength of conservative, dissipative, and random forces, respectively.
$\omega^D$ and $\omega^R$ are weight functions and $\xi_{ij}$ is a symmetric
normally-distributed random variable with zero mean and unit variance. All forces act
within a sphere of cutoff radius $r_c$. The random and dissipative forces form
a thermostat and have to satisfy the fluctuation-dissipation theorem \cite{Espanol1995}
to maintain the equilibrium temperature $T$ in DPD system. This results in two conditions,
\begin{equation}
\omega^D(r_{ij})  =  \left[ \omega^R(r_{ij})\right]^2, \quad \quad \sigma_{dpd}^2  =  2\gamma k_B T. 
\end{equation}
The typical choice for the weight function is $\omega^R(r_{ij})= (1-r_{ij}/r_c)^s$, where
$s$ is a selected exponent. 

Rigid walls are simulated by immobilized DPD particles having identical number density and interactions 
as the fluid particles. The immobile wall particles enforce no-slip boundary conditions (BCs) at the walls through 
dissipative interactions with the fluid particles. In order to prevent wall penetration by fluid particles, reflective surfaces 
are introduced at the interface between the fluid and the walls. Slip BCs at the walls can be modeled by turning off 
the dissipative interactions between fluid and wall particles. 

\noindent
{\bf Model of active particles and passive tracers in simulations.} Both active swimmers and passive tracers are represented by an arrangement of particles uniformly 
distributed on a spherical surface and forming a triangulated network (i.e. membrane model). To ensure the structural integrity of 
the triangulated network, it is supplied with shear elasticity, bending rigidity, and constraints related to the 
surface area and enclosed volume \cite{Fedosov_RBC_2010}. The bonds connecting adjacent vertices within 
the network provide shear elasticity defined by a potential function $U_e = U_{wlc} + U_{pow}$ as \cite{Fedosov_RBC_2010}.
\begin{equation}
	U_{wlc} = \frac{k_BT l_m}{4p}\frac{3x^2-2x^3}{1-x}, \quad \quad U_{pow} = \frac{k_p}{l},	
\end{equation}
where $l$ is the spring length, $l_m$ is the maximum spring extension, $x = l/l_m$, $p$ is the persistence length, 
and $k_p$ is the repulsive force coefficient. The bending rigidity is described by the Helfrich bending energy
\cite{Helfrich_EPB_1973}
\begin{equation}
	U_b= \frac{\kappa}{2} \sum_i \sigma_{i} (H_i-H_0^i)^2,
	\label{helfirch_eq_des}
\end{equation}
where $\kappa$ is the bending rigidity, $H_i = {\bf n}_{i} \cdot \sum_{j(i)} \sigma_{i j} {\bf r}_{i j}/(\sigma_{i}r_{i j})$ is the discretized mean curvature at
vertex $i$,  ${\bf n}_{i}$ is the unit normal at the membrane vertex $i$, $\sigma_{i} =\sum_{j(i)} \sigma_{i j}r_{i j}/4$ is the area corresponding
to vertex $i$ (the area of the dual cell), $j(i)$ corresponds to all vertices linked to vertex $i$, and $\sigma_{i j}=r_{ij}(\cot\theta_1+ \cot\theta_2)/2$
is the length of the bond in the dual lattice, where $\theta_1$ and $\theta_2$ are the angles at the two vertices opposite to the edge $ij$ in the dihedral.
Finally, $H_0^i$ is the spontaneous curvature at vertex $i$. 

The area and volume conservation constraints are given by \cite{Fedosov_RBC_2010}
\begin{equation}\label{eq:av}
	U_a = \frac{k_a(A-A_0^{tot})^2}{2A_0^{tot}} + \sum_{m\in 1...N_t}{\frac{k_d(A_m-A^m_0)^2}{2A^m_0}},
\end{equation}
\begin{equation}  
	U_v =  \frac{k_v(V-V_0^{tot})^2}{2V_0^{tot}},
\end{equation}
where $A$ is the current area of the membrane, $A_0^{tot}$ is the targeted global area, $A_m$ is the area of the m-th triangle from $N_t$ triangles 
on the triangulated surface, $A^m_0$ is the targeted area of the m-th triangle determined by the triangular mesh on the particle surface, $V$ is 
the instantaneous membrane volume, and $V_0^{tot}$ is the targeted volume. Furthermore, $k_a$, $k_d$, and $k_v$ are the global
area, local area and volume constraint coefficients, respectively.

Both swimmers and tracers are submerged within a DPD fluid, and also filled by fluid DPD particles due to their membrane-like representation. The membrane surfaces
of all suspended active particles and tracers serve as a boundary separating DPD particles inside and outside of the membranes. This is achieved through the reflection 
of solvent particles at the membrane surfaces from inside and outside. Note that the dissipative and random forces between the internal and external fluid particles are 
deactivated, and only the conservative force is employed to maintain uniform fluid pressure across the membranes. No slip BCs between the fluid and 
suspended colloidal particles are implemented through the proper adjustment of the friction coefficient $\gamma$ between fluid and membrane particles \cite{Fedosov_RBC_2010}.    

Propulsion of active swimmers is realized through a slip velocity at their surface using the spherical squirmer model \cite{Alarcon_MCP_2017} 
\begin{equation}
	\label{eq:surface_velocity}
	{\bf u}_s = \left( B_1 \sin(\phi) + B_2 \sin(\phi)\cos(\phi)\right) {\bf t_{\phi}},
\end{equation}
where ${\bf t_{\phi}}$ represents the local tangential vector in the $\phi$ direction at the swimmer surface, and $\phi$ corresponds to the angle between the orientation 
of the squirmer and the vector extending from its center of mass to a local position on its surface. The coefficient $B_1$ governs the propulsion strength of the squirmer, 
while $B_2$ introduces a front-back asymmetry into the slip-velocity field. In bulk, the spherical squirmer swims with a speed $v_0 = 2 B_1 /3$. We employ the dimensionless 
ratio $\beta = \frac{B_2}{B_1}$ to characterize the mode of propulsion, where $\beta < 0$ represents a pusher, $\beta = 0$ corresponds to a neutral swimmer, and $\beta > 0$ 
models a puller. To enforce the slip velocity at the squirmer surface, the dissipative interaction between the squirmer particles and those of the surrounding fluid is adjusted as follows
 \begin{equation}
	{\bf F}_{ij}^D = -\gamma \omega^D(r_{ij})(\hat{\bf r}_{ij}\cdot{\bf v}_{ij}^\ast)\hat{\bf r}_{ij}, \quad  \quad 
	{\bf v}_{ij}^\ast = {\bf v}_i - {\bf v}_j + {\bf u}_s^i,
\end{equation}
where ${\bf u}_s^i$ is the slip velocity at the position of squirmer particle $i$, while $j$ corresponds to an outer-fluid DPD particle. 

Excluded volume interactions between active squirmers and tracers are imposed through the 12-6 Lennard-Jones (LJ) potential as
\begin{equation}
	U_{LJ}(r)= 4\epsilon \left[\left(\frac{\sigma_{LJ}}{r}\right)^{12}- \left(\frac{\sigma_{LJ}}{r}\right)^{6}\right], 
	\label{eq:LJ}
\end{equation}
where $\sigma_{LJ}$ is the distance at which $U_{LJ}(\sigma_{LJ})=0$ and $\epsilon$ is the potential strength. Note that only the repulsive part of the LJ potential is employed 
by setting the cutoff radius to $r_c^{LJ}=2^{1/6}\sigma_{LJ}$.

\noindent
{\bf Simulation setup and parameters.} The simulation domain corresponds to a slit-like geometry with a size of $100r_c \times 100r_c  \times 22r_c$ and periodic BCs in the xy plane. In the $z$ axis, the 
fluid layer of thickness $20r_c$ is confined between two walls, each with a thickness of $r_c$. The fluid modeled by a collection of DPD particles has a number density 
of $n=5/r_c^3$ ($r_c=1.0$ is selected in simulations). The DPD force coefficients for interactions between fluid particles are $b=50 k_B T / r_c$, $\gamma=14.15 \sqrt{m k_B T}/ r_c$, 
and $s=0.15$, where $m=1$ is the particle mass and $k_B T = 2$ is the unit of energy. These parameters lead to a fluid dynamic viscosity of $\eta=16.7 \sqrt{m k_B T}/ r_c^2$. 

Suspended squirmers are constructed from $1000$ particles placed on a spherical surface with a diameter $d_{Sq} = 6.4 r_c$. The largest investigated tracers have the same diameter. 
The area fraction of $\Phi=0.05$ corresponds to $18$ simulated squirmers, to which $10$ tracers were added. Both the squirmers and tracers are placed at the lower wall ($z=0$) with 
no-slip BCs, and restricted to a fluid layer of thickness $7.3r_c$, in order to properly capture the quasi-2D environment in experiments. At the upper wall ($z=20r_c$) slip BCs are 
imposed, which partially accounts for a larger domain in experiments. Coupling between the squirmer/tracer particles and fluid particles assumes $b=0$ and $\gamma=12 
\sqrt{m k_B T}/ r_c$ with $s=0.1$ and $r'_c=0.7 r_c$. The value of $\gamma$ here is computed such that the imposed slip BCs at the squirmer surface are obtained \cite{Fedosov_RBC_2010}. 
The time step in simulations is $\Delta t = 0.007 r_c \sqrt{m / k_B T}$. 

To relate simulation and physical units, we introduce characteristic length and time scales. The length scale is based on the squirmer diameter $d_{Sq} = 6.4 r_c$, resulting in 
$r_c$ to represent approximately $0.34$ $\mu m$ ($d_S = 2.16$ $\mu m$ in experiments). Therefore, the simulation domain corresponds to $34 \times 34 \times 6.8$ $\mu m$. 
The time scale is $\tau = d_{Sq}/v_M$, where $v_M$ is the modal speed of the speed distribution of active particles in the xy plane, see Fig.~1(b) in the main text.   

\begin{figure}
\centering
   \includegraphics[width=0.9\linewidth]{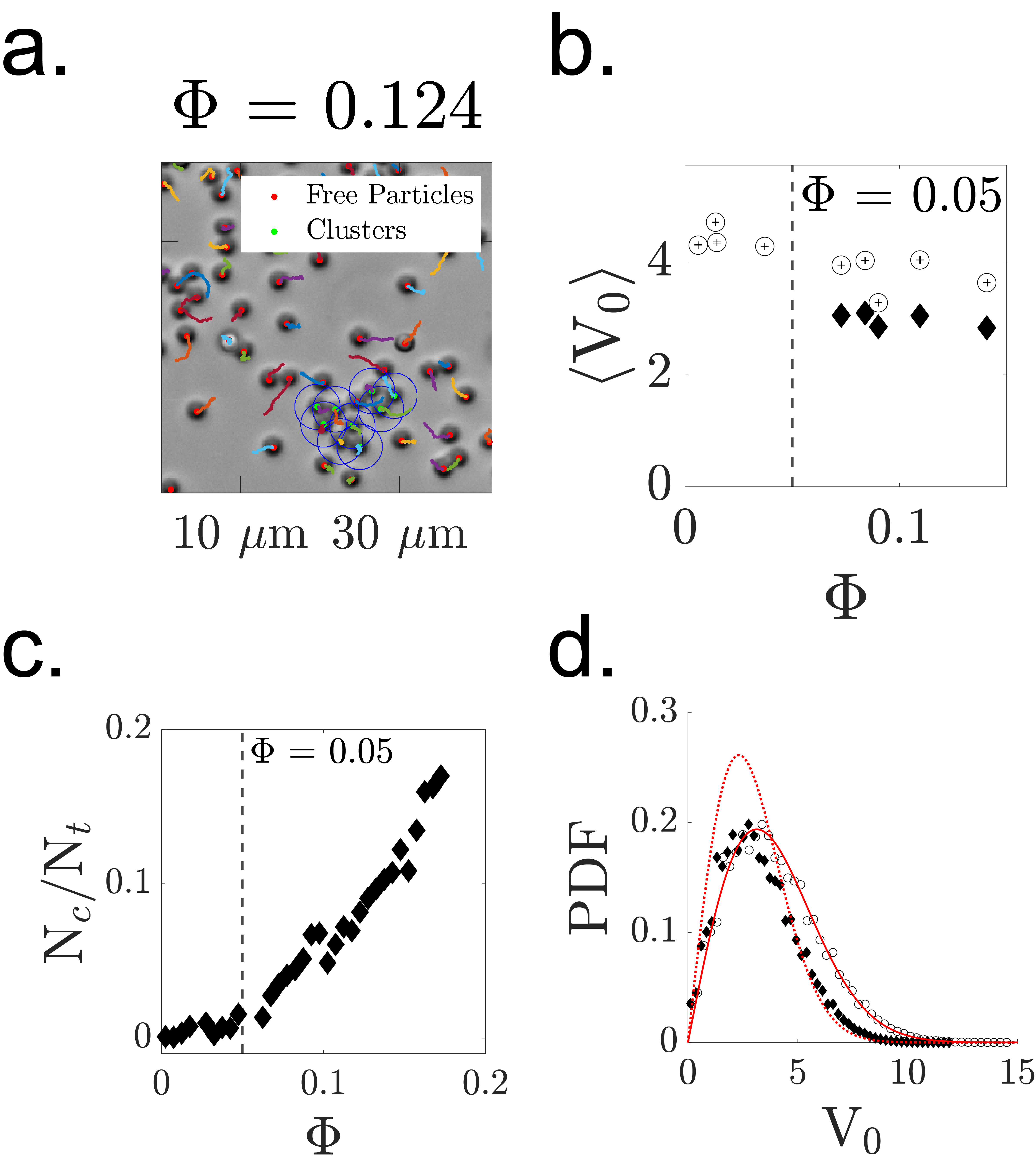} 
  \caption{Separating “clustered" particles from “free" particles. a. Identification of “clustered" particles from micrographs. Clusters are defined as particles with 2 or more nearest neighbours within 2.5R for at least 12 frames (0.24 s, \cite{Theurkauff2012}). b. Mean speed of free (open circle) and clustered (closed diamond) microswimmers. c. Onset of clustering at an area fraction $\Phi \sim 0.05$. d. Instantaneous speed distributions of free and clustered particles (symbols as before). The red curves (solid and dashed for the free and clustered particles respectively) show the best fits of the Maxwell distribution \cite{Wu2000}. We clearly see that clustered particles cannot be well explained by this distribution, while the motion of the non-interacting microswimmers is well captured. Notably, the distribution of speeds follows that for neutral squirmers (see main text, Figure 4).}
   \label{suppfig:veldistbn} 
\end{figure}

\begin{figure}
\centering
   \includegraphics[width=0.9\linewidth]{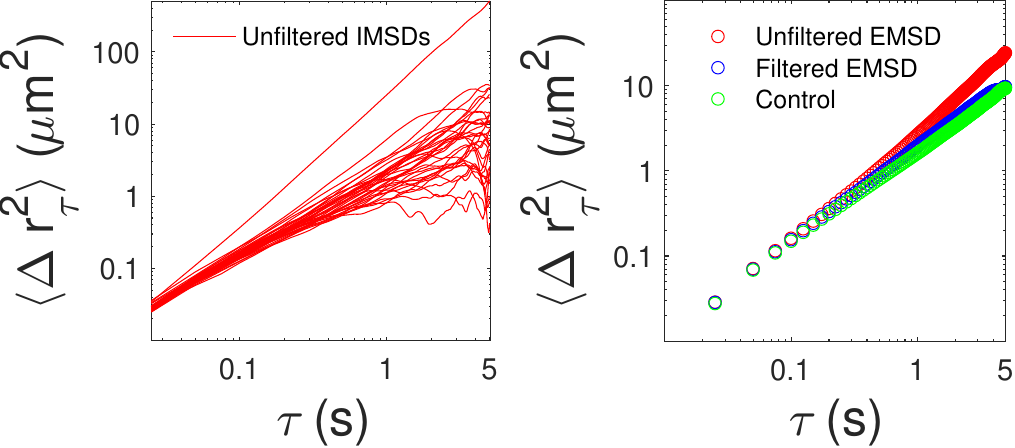} 
  \caption{Effect of “actively transported" colloids on the ensemble-mean-squared-displacement (EMSD) of all tracers. Left: The individual MSDs (IMSDs) of multiple tracer particles (here, $1 \mu m$) in a dilute suspension of synthetic microswimmers. We see one extreme outlier, which corresponds to a singular tracer particle being actively transported by a microswimmer (via phoretic attraction). Right: The single actively transported tracer significantly contributes to the overall EMSD (red - due to averaging of all values). Removing this single IMSD from the ensemble (blue) collapses the EMSD of the tracers onto the control EMSD (green) of passive tracers in the absence of active microswimmers. This highlights the issues of blindly using averaged values alone when evaluating systems with large outliers.}
   \label{suppfig:EMSDS_tracer}
\end{figure}

\begin{figure}
\centering
   \includegraphics[width=0.9\linewidth]{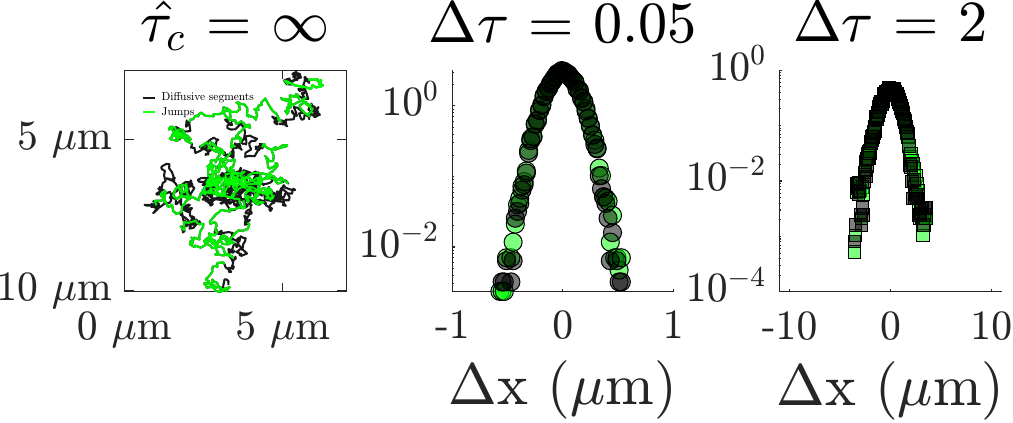} 
  \caption{Tracer displacements in the presence of microswimmers which are passively diffusing (no UV illumination). At both short and long lag times $\Delta\tau$, the distribution is Gaussian. Contrast this to the enhanced diffusivity of tracers in the presence of motile microswimmers, shown in Figure 3 (top row) of the main text. Note that the autocorrelation approach to extracting jumps also identifies jumps in the passive case, but that the resulting distributions are unaffected. 
  }
   \label{suppfig:passive_distbn} 
\end{figure}

\begin{figure}
\centering
   \includegraphics[width=\linewidth]{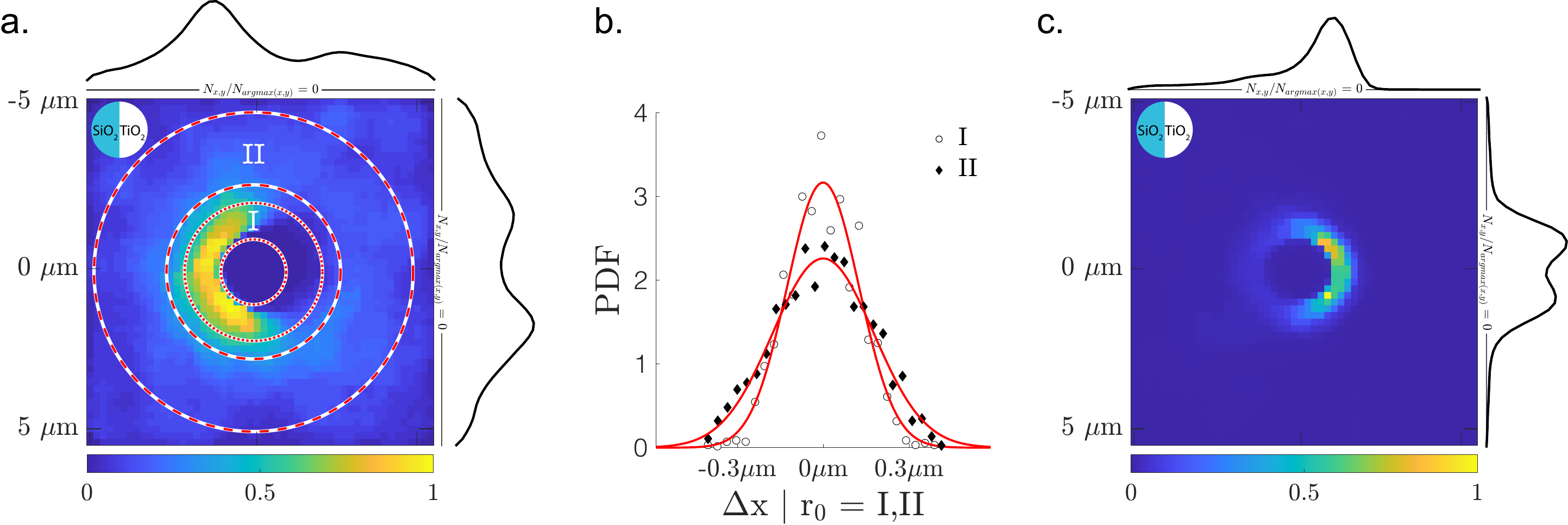} 
  \caption{Demonstration of the phoretic interactions between the Janus microswimmers and the PS tracer nanoparticles. By immobilizing our microswimmers and modulating the pH, we also observe a phoretic contribution to the generated flow fields, in agreement with previous reports for TiO\textsubscript{2}-SiO\textsubscript{2} systems \cite{Singh2017}. The values are averaged over at least 4 Janus pumps (stuck to the substrate via drying) in each case. Specifically, we note the change in the accumulation of the tracer nanoparticles as tetramethylammonium hydroxide (TMAH) is added to modulate the pH, following \cite{Singh2017}. a. Accumulation of PS nano-tracers around a fixed Janus particle at the SiO\textsubscript{2}  hemisphere under standard experimental conditions in the presence of H\textsubscript{2}O\textsubscript{2} with UV illumination (pH $\sim 4$, orientation shown in inset, top left). Colour coding indicates the normalised number density, as well as the averaged profiles in X and Y. b. Distribution of displacements of the nanoparticle tracers in the regions marked I and II in a. We see that near the Janus particle, tracer displacements decrease due to accumulation. Far from the particle, we observe the expected Gaussian distribution (also see Figure S5). c. Accumulation of PS nano-tracers around a fixed Janus particle at the TiO\textsubscript{2}  hemisphere under modulation of the pH (pH $\sim 7$) via addition of TMAH, in the presence of H\textsubscript{2}O\textsubscript{2} with UV illumination (orientation shown in inset, top left). For these experiments, tracers were visualised using a 100x oil objective.} 
   \label{suppfig:Phoretic} 
\end{figure}

\begin{figure}
\centering
   \includegraphics[width=0.9\linewidth]{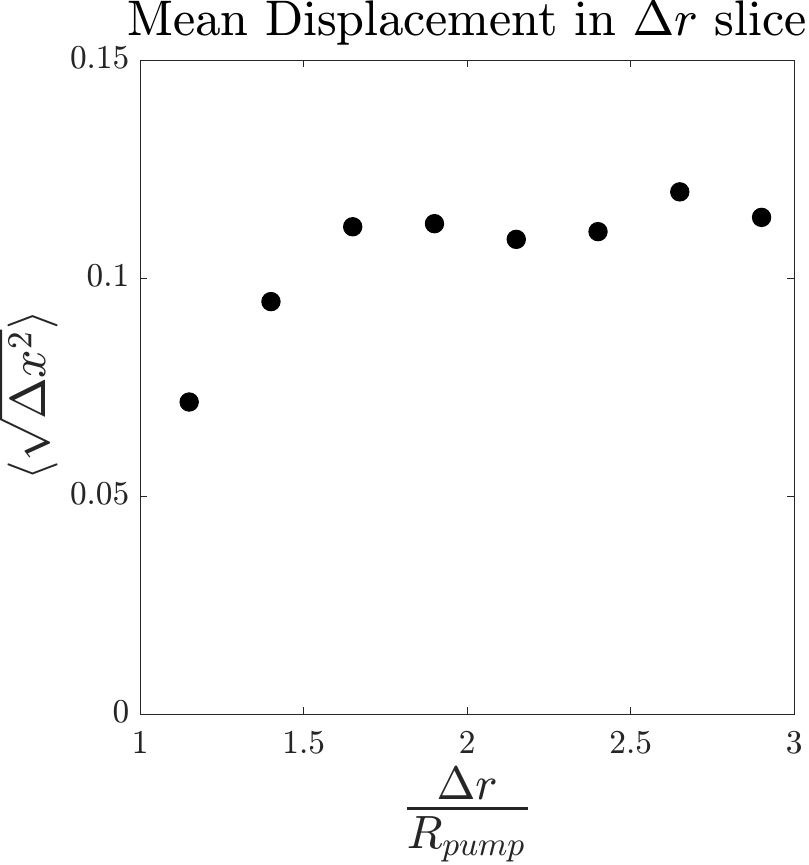} 
  \caption{Average instantaneous displacement of tracer nanoparticles in rings evaluated around the Janus pumps evaluated in Figure S4 a. We see that outside of the inner rings (where the tracers accumulate due to phoretic interactions), the average displacements of the tracers rapidly saturate, indicating little-to-no effect of the far-field hydrodynamic flows created by the microswimmer on the tracer dynamics}
   \label{suppfig:pump}
\end{figure}

\begin{figure}
\centering
   \includegraphics[width=\linewidth]{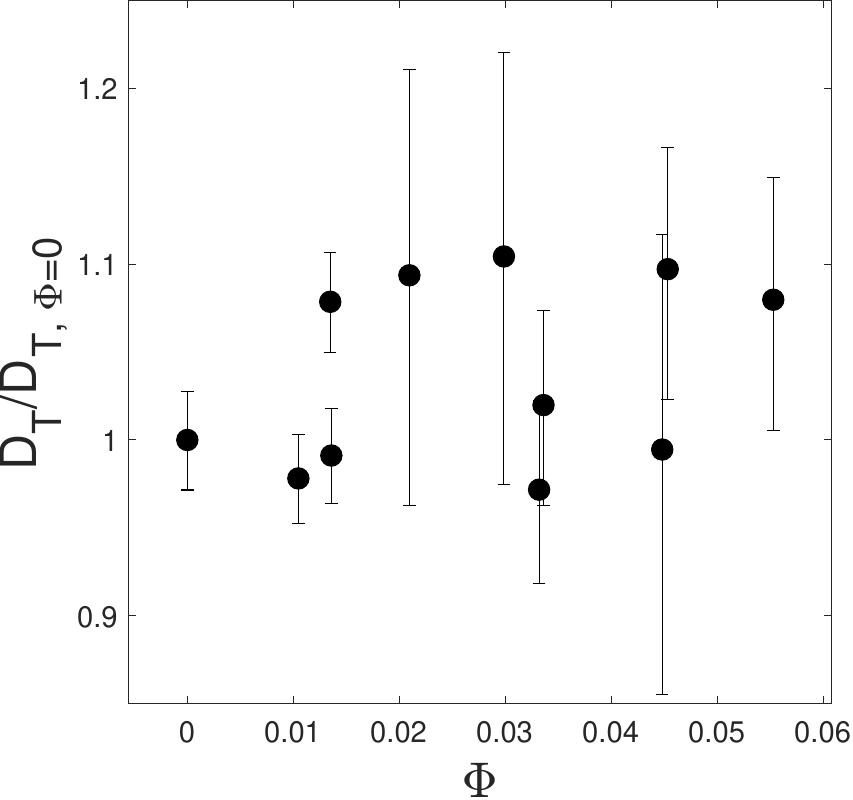} 
  \caption{Fitted translational diffusion coefficients $D_T$ of tracer nanoparticles in the presence of different concentrations (represented by area fraction, $\Phi$) of microswimmers, normalised by the control at $\Phi = 0$. The mean value and error bars are obtained from the distribution of $D_T$ fitted from the mean logarithmic squared displacement\cite{Kepten2013,Bailey2022_PRE} (MLSD, given by $MLSD = log(4D_T) + \alpha log(\tau)$), assuming a log-normal distribution (also see Figure S7 c.). We see that there is no clear trend in the values of $D_T$ with the presence of microswimmers, indicating the lack of an enhancement in their diffusivity (via mixing), which we attribute to their high initial diffusion coefficient (arising from their small size).
  }
   \label{suppfig:DT_NPs} 
\end{figure}

\begin{figure}
\centering
   \includegraphics[width=0.9\linewidth]{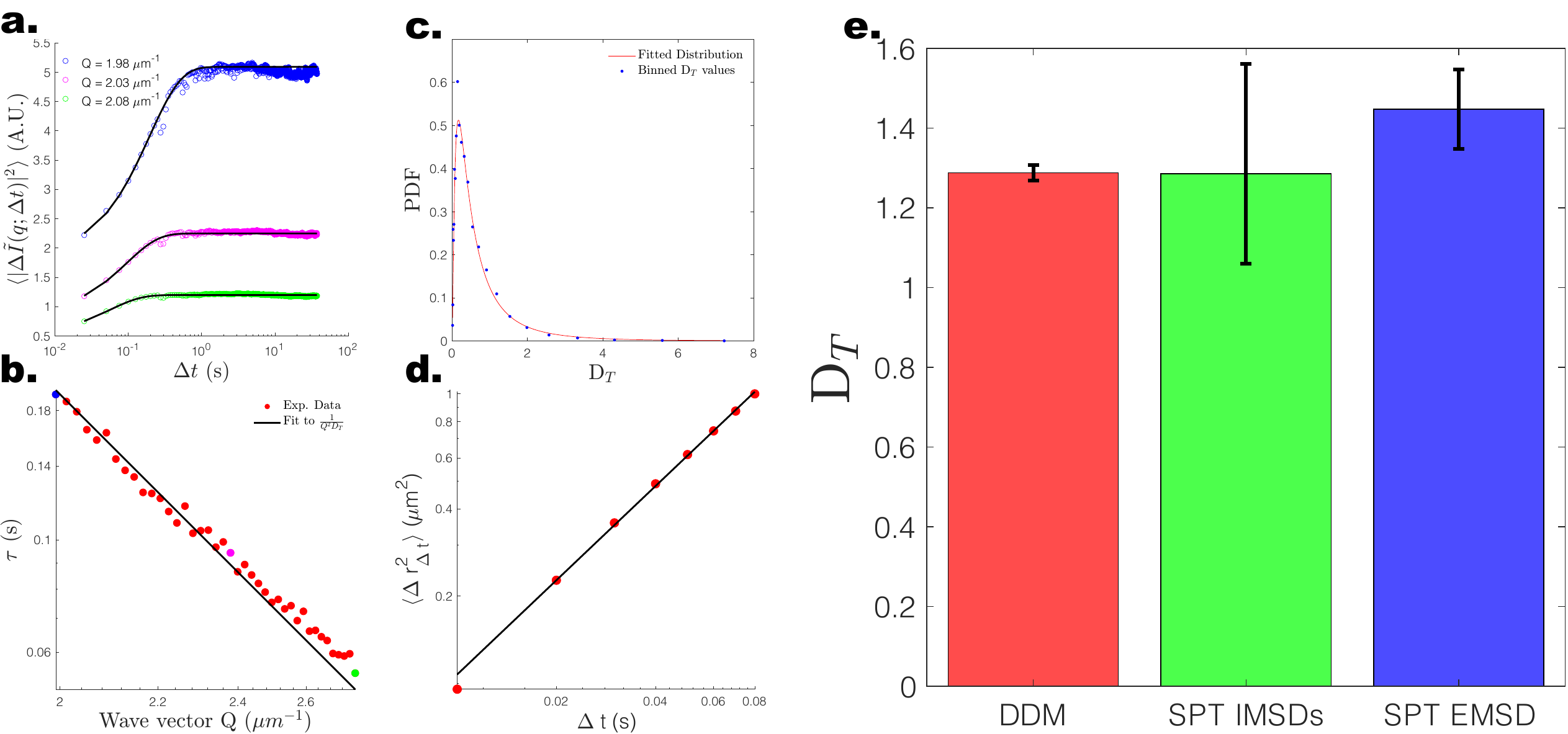} 
  \caption{Comparison of the fitting approaches used to extract $D_T$ values of the (control, i.e. in the absence of microswimmers) PS nanoparticle tracers, to confirm the accuracy of various approaches. a-b. Fitting of $D_T$ via differential dynamic microscopy (DDM \cite{Cerbino2008}). c. Fitting a log-normal distribution to the binned values of $D_T$ obtained using the MLSD of individual tracers (also see Figure S6) d. Fitting the EMSD of the tracers directly. e. Comparison of the fits obtained using different approaches (in $\mu m^2s^{-1}$). We see good agreement between using DDM and the distribution of the MLSDs of individual nanoparticle tracers in the passive case. As the presence of microswimmers (additional scattering sources) affects the interpretation of DDM results, we therefore use the values obtained from the distribution of the IMSDs.
  }
   \label{suppfig:ComparisonFitting} 
\end{figure}

\end{document}